# Research Landscape of the novel emerging field of Cryptoeconomics

Alican Alaşık*, Nihan Yıldırım†

**Abstract.** A bibliometric literature analysis was conducted to illuminate the evolving and rapidly expanding literature in the field of cryptoeconomics. This analysis presented the emerging field's intellectual, social, and conceptual structure. The intellectual structure, characterized by schools of thought, emerged through a common citation analysis. The social structure revealed collaborations among researchers, identified through a co-authorship analysis. Network analysis highlighted collaborative communities facilitating innovation and knowledge exchange within the field. The conceptual structure was enlightened by analyzing common terms occurring in titles, author keywords, abstracts, and the publication itself. This bibliometric analysis of the rapidly advancing field of cryptoeconomics serves as a foundational resource, providing insights into research productivity and emerging trends. It contributes to a deeper understanding of the field, offering valuable information on research patterns and trends. Furthermore, this analysis empowers researchers, policymakers, and industry sectors to make informed decisions, establish collaborations, and navigate the dynamic and evolving landscape of the cryptoeconomics field.

## 1. Introduction

Cryptoeconomics is a novel field that emerged following the discovery of distributed ledger technology (DLT). Vlad Zamfir (2015) of the Ethereum developer community is the person who made the earliest recorded definition[1] of cryptoeconomics: *"... a formal discipline that studies protocols that govern the production, distribution, and consumption of goods and services in a decentralized digital economy. Cryptoeconomics is a practical science that focuses on designing and characterizing these protocols*". Cryptoeconomics can disrupt traditional institutions by enabling decentralized and trustless systems[2], and introducing new economic models[3]. Cryptoeconomics explores the design and analysis of decentralized systems, particularly those based on DLT, by integrating cryptographic techniques and economic incentives. At its core, cryptoeconomics aims to create robust and secure systems that incentivize desirable behaviour among participants without relying on a central authority. Cryptoeconomic systems (CESs) provide new kinds of economic coordination and new forms of organizational design[4] [5]to share resources[6] [7]. Cryptoeconomic networks are techno-socio-economic complex systems[8] [9] *defined by (i) individual autonomous actors, (ii) economic policies embedded in software (the protocol or smart contract code), and (iii) emergent*

*A. Alaşık (alasik19@itu.edu.tr) is an M.Sc. student at the Department of Management Engineering, Istanbul Technical University
†N. Yıldırım (yildirimni@itu.edu.tr) is an Associate Professor at the Department of Management Engineering, Istanbul Technical University

*properties arising from the interactions of those actors with the whole network, according to the rules defined by that software*[10].

Distributed ledger technology and the emerging cryptoeconomic systems that use it have created new chances to respond to grand societal challenges humanity faces, such as sustainability and climate change[11]. However, cryptoeconomics is a multidisciplinary and still evolving field[12] [13] that isn't easy for many researchers to understand due to its complex, interdisciplinary, and technical nature[14]. Science mapping is increasingly becoming a must-do activity for scientists of many disciplines[15]. There are bibliometric analyses on blockchain technology[16] [17] [18] [19] [20], but none on cryptoeconomics. Therefore, we conducted a bibliometric analysis to shed light on the emerging and rapidly growing field of cryptoeconomics. In this bibliometric study, while examining the field or conceptual structure of cryptoeconomics, we will determine the field's knowledge base and intellectual structure and show the scientific community's social network structure.

## 2. Method

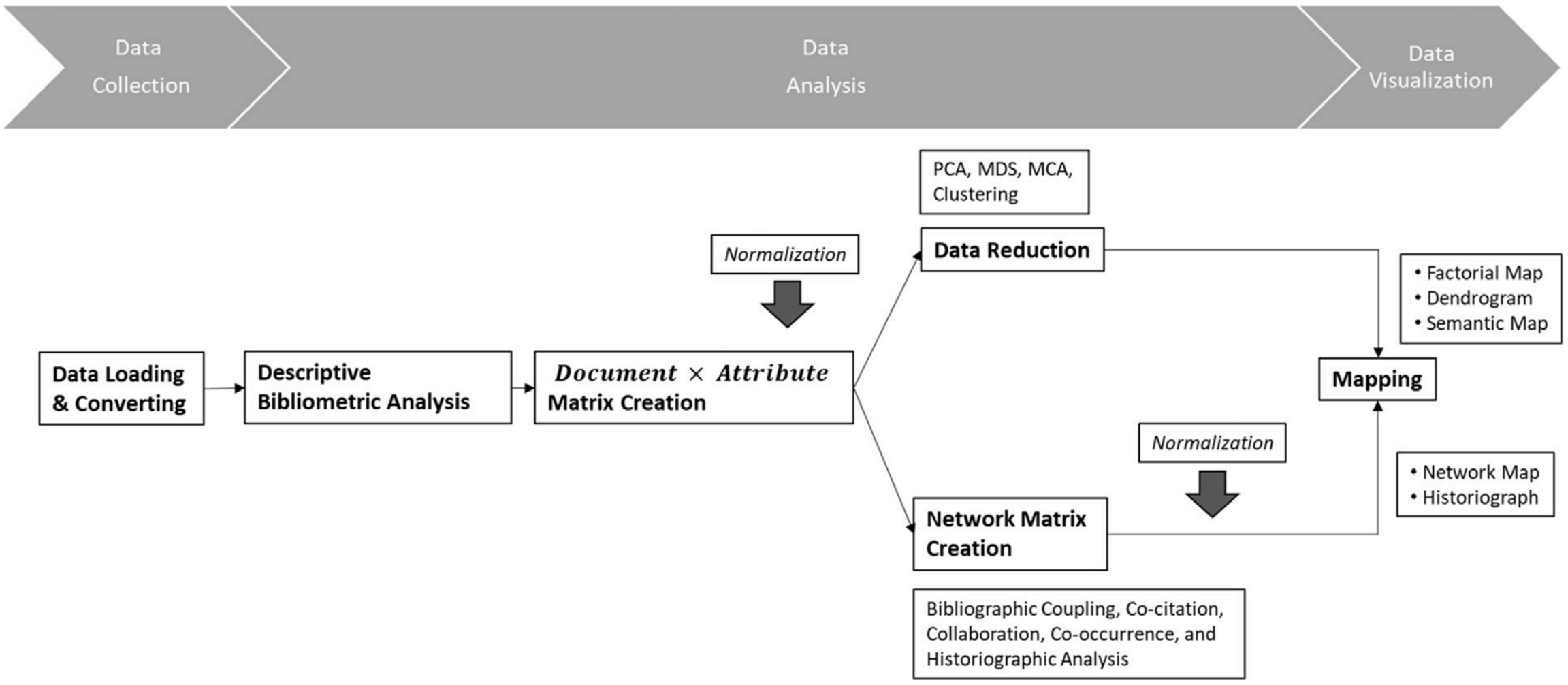

Fig. 1. Science Mapping Workflow (Aria & Cuccurullo, 2017).

In this bibliometric study, we will examine the field or conceptual structure of cryptoeconomics, determine the field's knowledge base and intellectual structure, and show the scientific community's social network structure. While designing this study, we divided the science mapping workflow (Fig. 1) into three parts: data collection, data analysis, and data visualization.

*Data collection*— Cryptoeconomics system design was first implemented by Nakamoto in "Bitcoin: A Peer-to-Peer Electronic Cash System"[21] and conceptually defined[22] by the Ethereum core developer team in 2015. However, the first academic publication on the topic was made in 2017. The data was gathered manually by entering the search query into the Scopus database. The data was processed using filters to exclude irrelevant information from the search results (such as subject fields, document formats, and language). Then, PRISMA[23] protocols developed by Liberti et al. were applied to improve and maintain the accuracy of review studies. Searching for keywords in the Scopus database source resulted in 179 documents. After, the documents were scanned and filtered for English language, sectoral research, and document type, yielding 171 papers between 2017 and 2023 (September).

*Data analysis*—Network analysis enables us to perform statistical analysis on the obtained maps to indicate different measures of the overall network or measurements of the relation or overlap of the distinct clusters determined[24]. The attributes of a document are linked to one another via the document itself. A Document × Attribute matrix can represent the links between distinct attributes. An attribute is a piece of data connected with a document recorded in a field tag within the bibliometric data frame (citation, author, and word.)

Citation analysis is the most popular type of bibliometric analysis[25]. It uses citation counts to calculate the similarity of documents, authors, and journals to explore intellectual structure[26]. In this bibliometric study, the analysis of co-citations focused on the author,[27] and this gave us the schools of thought of the theories that developed around cryptoeconomics. A co-citation network can be obtained using the general equation (1):

$$B_{cocit} = A \times A' \tag{1}$$

where A is a Document × Cited reference matrix. Element b$ij$ indicates how many co-citations exist between documents $i$ and $j$. The main diagonal of $B_{cocit}$ contains the number of documents where a reference is cited in our data frame.

A scientific collaboration network is one in which nodes represent authors and linkages represent co-authorships. The writers and their affiliations are examined in co-author analysis to explore the social structure and collaborative networks[28]. An author collaboration network can be obtained using the equation (2):

$$B_{coll} = A \times A' \tag{2}$$

where A is a Document × Author matrix. Element b$ij$ indicates the number of collaborations between authors $i$ and $j$. The diagonal element b$ii$ is the number of documents authored or co-authored by researcher $i$.

Co-word analysis studies the conceptual structure of the research field by analyzing the most important words or keywords in papers[29]. This analysis aims to create the conceptual structure of a framework by mapping and clustering terms derived from keywords, titles, or abstracts in a bibliographic collection using a word co-occurrence network[30]. It can also detect a research field's conceptual structure, exposing thematic clusters and their relationships[31]. It is the only method that constructs a similarity measure based on the actual content of the documents; the others relate documents indirectly through citations[32]. A co-word network can be obtained using the general equation (3):

$$B_{coc} = A \times A' \tag{3}$$

where A is a Document × Word matrix, where Word is, alternatively, authors' keywords, keywords plus, or terms extracted from titles or abstracts. Element b*ij* indicates how many co-occurrences exist between words *i* and *j*. The diagonal element b*ii* is the number of documents containing the word *i*.

*Data Visualization*—The next section presents it using the Bibliometrix R package under three subheadings: intellectual, social, and conceptual structure.

## 3. Science Mapping

*Intellectual Structure*— In the co-citation analysis, it is not surprising to see that the intensity is concentrated on Nakamoto, followed by Vitalik Buterin (Fig. 2.) Nakamoto published his (Bitcoin: A Peer-to-Peer Electronic Cash System) article in 2008, giving rise to the field and Buterin's (2014) addition of *smart contracts*[33] to this cryptoeconomic system idea added dimension to the cryptoeconomics field. Although the smart contract idea started to be discussed in the 80s, Nick Szabo[34] who used this name for the first time in 1996.

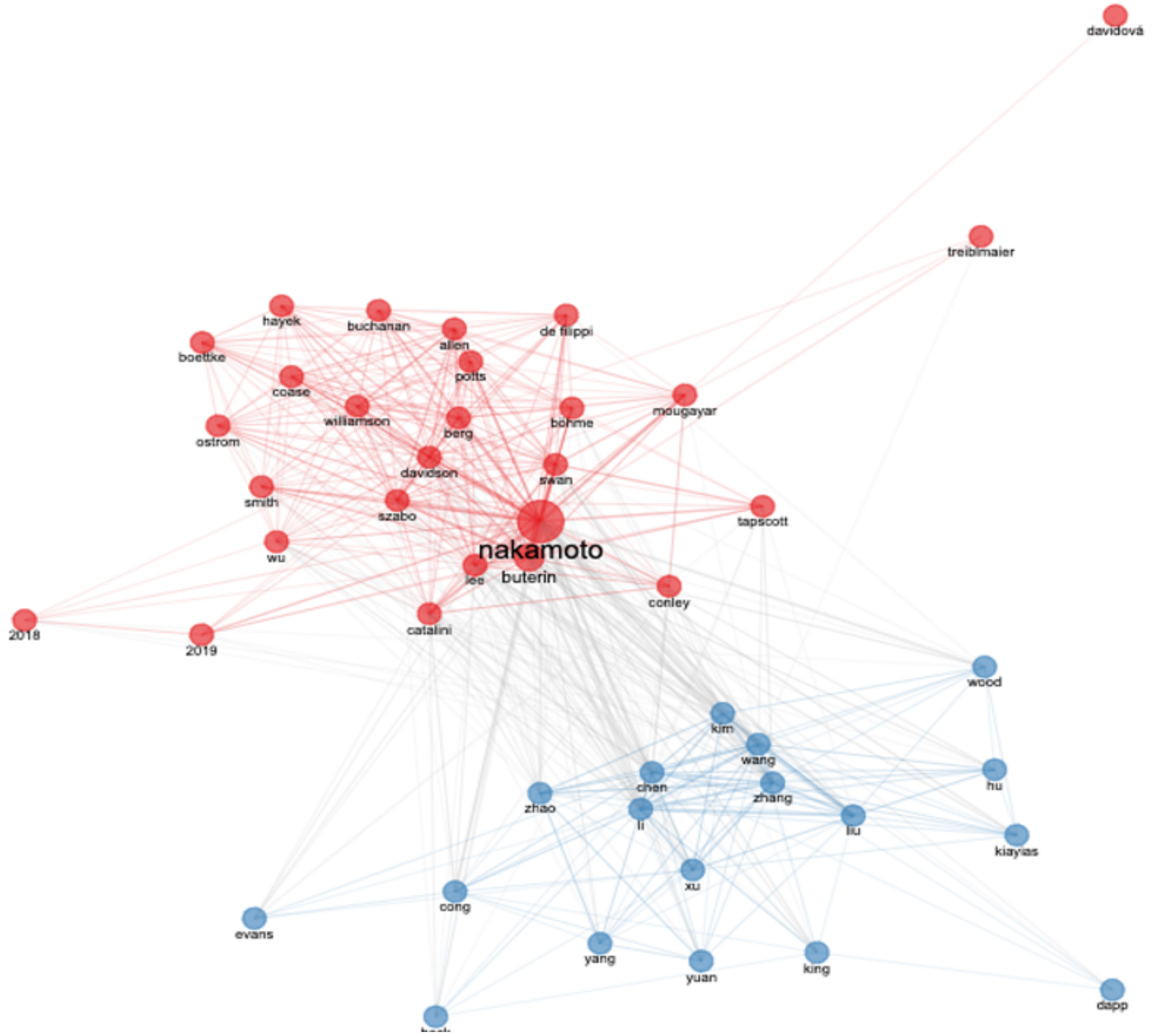

Fig. 2. Co-citation Network (author)

We can also see that there are predominantly two clusters that are citing each other (each color represents one cluster in Fig 2). While the majority of the red cluster in the co-citation network consists of writers working in economics, especially in the institutional field, it seems that the blue cluster is concentrated on the informational side. If an economic analysis studies social institutions coordinating and governing transactions, it can be called institutional[35].

To uncover and explore the intellectual structure of cryptoeconomics field, the prominent authors and their works in Fig 1 were carefully examined. Friedrich Hayek[36] (1945) discovered the role of dispersed information and its connection to prices. James M. Buchanan[37] (1965) developed club theory. Ronald Coase[38] (1937) studied transaction costs and explained the nature of firms. Furthermore, Coase's transaction costs approach is currently essential to modern organizational economics, where it was reintroduced by Oliver E. Williamson[39]. According to Oliver Williamson, many of our economic, social, and political institutions have evolved to protect against opportunism by giving ex-ante or ex-post compensation for the uncertainty that rules will be followed[40]. The incomplete contracts approach to firm theory and corporate finance is partially based on the work of Williamson and Coase[41]. Ostrom's[42] (1990) work highlighted the role of public choice in decisions influencing the supply of public goods and services, and her work revealed the commons as an economic coordination mechanism. Berg, Potts, Allen, and Davidson who working on institutional cryptoeconomics. Institutional cryptoeconomics approaches blockchain as a new institutional form. Blockchain is an economic infrastructure that provides coordination and exchange for markets, firms, governments, clubs, and the commons[43].

Catalini identifies two significant costs impacted by distributed ledger technology: verification and networking costs[44]. He is also the founder of the MIT Cryptoeconomics Lab. Beck et al.[45] (2018) focus on theorizing in information systems (IS) research in the blockchain context. They offer a research agenda in the dimensions of decision rights, accountability, and incentives. Dapp's study (2019) is the first to introduce cryptoeconomics in the context of sustainability[46].

Co-citation analysis shows us that the intellectual structure of cryptroeconomics is divided into two clusters: Institutional and informational science. While political and economic theories are concentrated on the institutional side, on the informational side, computer science, complex systems science, and information systems explore cryptoeconomics.

*Social Structure*—We see which authors work together in the collaboration network (Fig. 3) we created through co-authorship. The thickness of the connections is proportional to the number of publications co-authored by the duo. For example, in Fig. 3, the thickest connection is between Potts J. and Berg C., as they are the two authors with the most joint publications, with six co-authored papers.

We conducted a detailed examination of the collaboration network, and authors have different collaborations. Seneviratne, O. and Godage, I. have collaborated inter-institutionally and interdisciplinary (Robotics and Blockchain)[47 48]. In Fig 3, among the authors in the red cluster, one of the three quintets with Zhang M. at the center, authors come from different institutions and disciplines (Computer Science and Business). However, there is no international collaboration between both groups; the first group is from the US, while the second group is from China. In the collaboration network, many groups collaborate within the same country, while the authors in the quintet with Peschanenko, V. come from the Netherlands and Ukraine. Still, they all belong to a similar discipline.

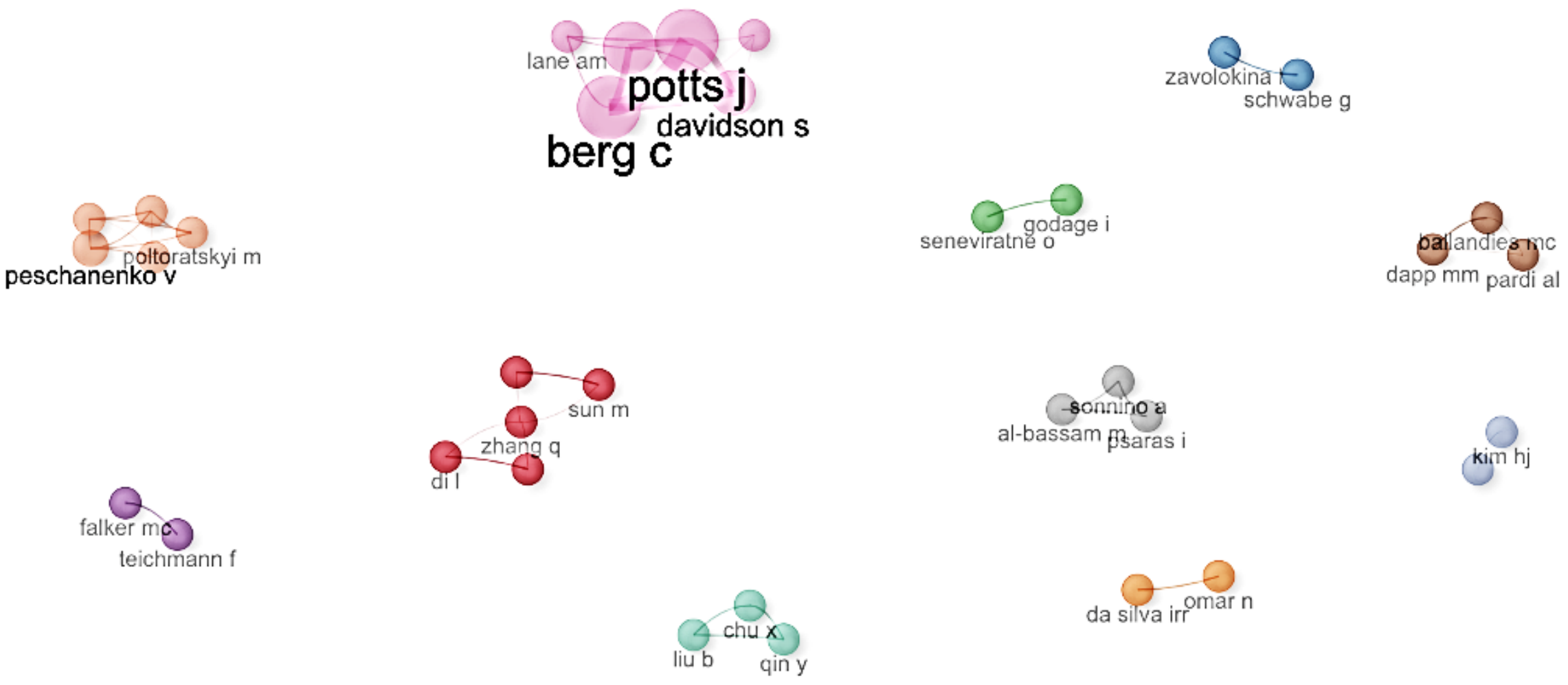

Fig. 3. Collaboration Network (author)

Over the years, International Research Collaboration (IRC) has grown in importance in assessing the scientific performance of individuals, organizations, and countries[49]. International collaborations in the field of cryptoeconomics are shown in Fig. 4. below.

In the international collaboration network (Fig. 4.), clustering is concentrated into two distinct groups, represented by red and purple. Among these countries, the UK, the US, and China are the most prominent in terms of collaboration. At the same time, the US and China are the countries that collaborate the most.

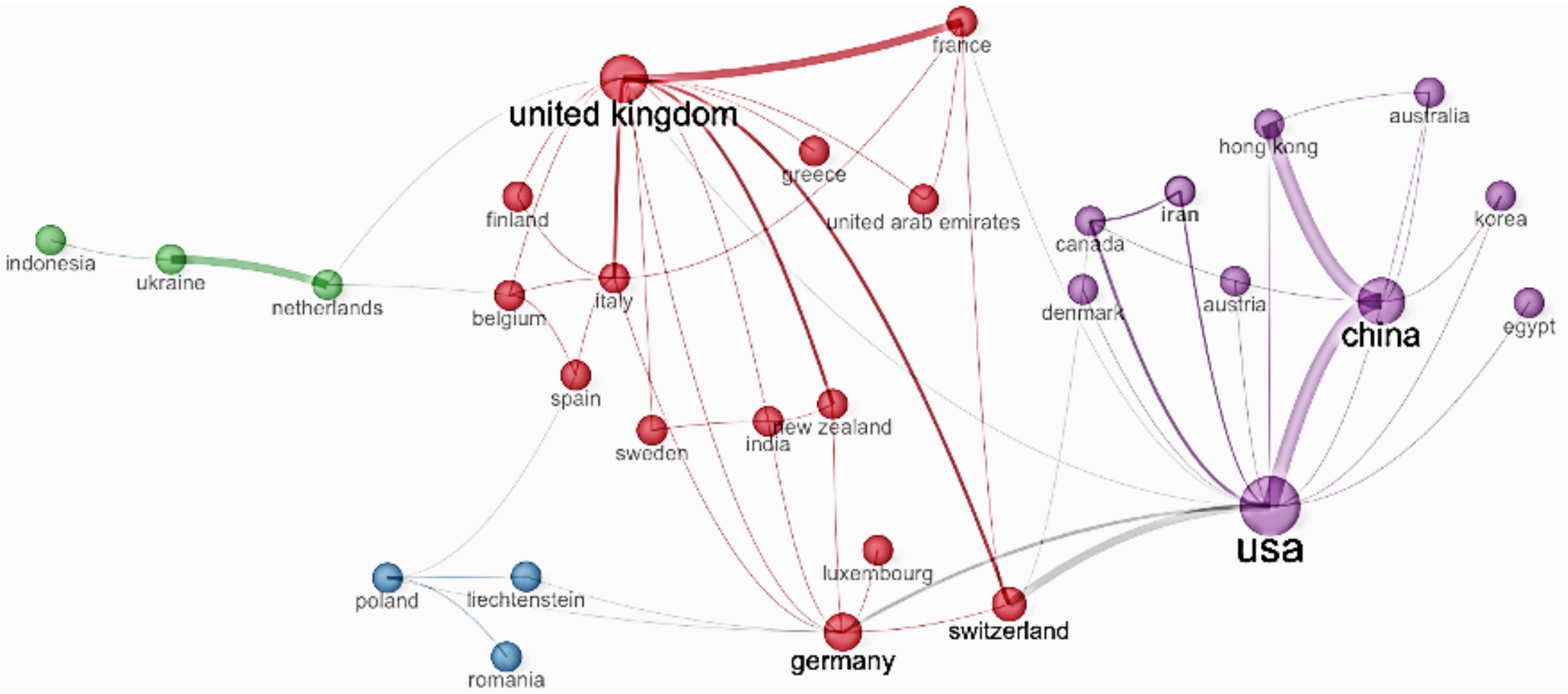

Fig. 4. International Collaboration Network

*Conceptual Structure*— We presented four co-occurrence networks, including keyword plus (Fig. 5.), author's keywords (Fig. 6.), title (Fig. 7.), and abstract (Fig. 8.). These networks can reveal patterns, structures, and dynamics that may not be apparent through other analytical approaches. Afterward, we listed disparate words (Table 1) and examined common (Table 2) among the Co-occurrence networks. This provides insights into relationships and interactions between concepts, allowing researchers and practitioners to understand better their relationships, dependencies, and potential functional connections.

First, keyword plus co-occurrence network (Fig. 5.) refers to additional keywords or terms extracted from the full text of articles. These are not the author's assigned keywords but are generated based on the article's content.

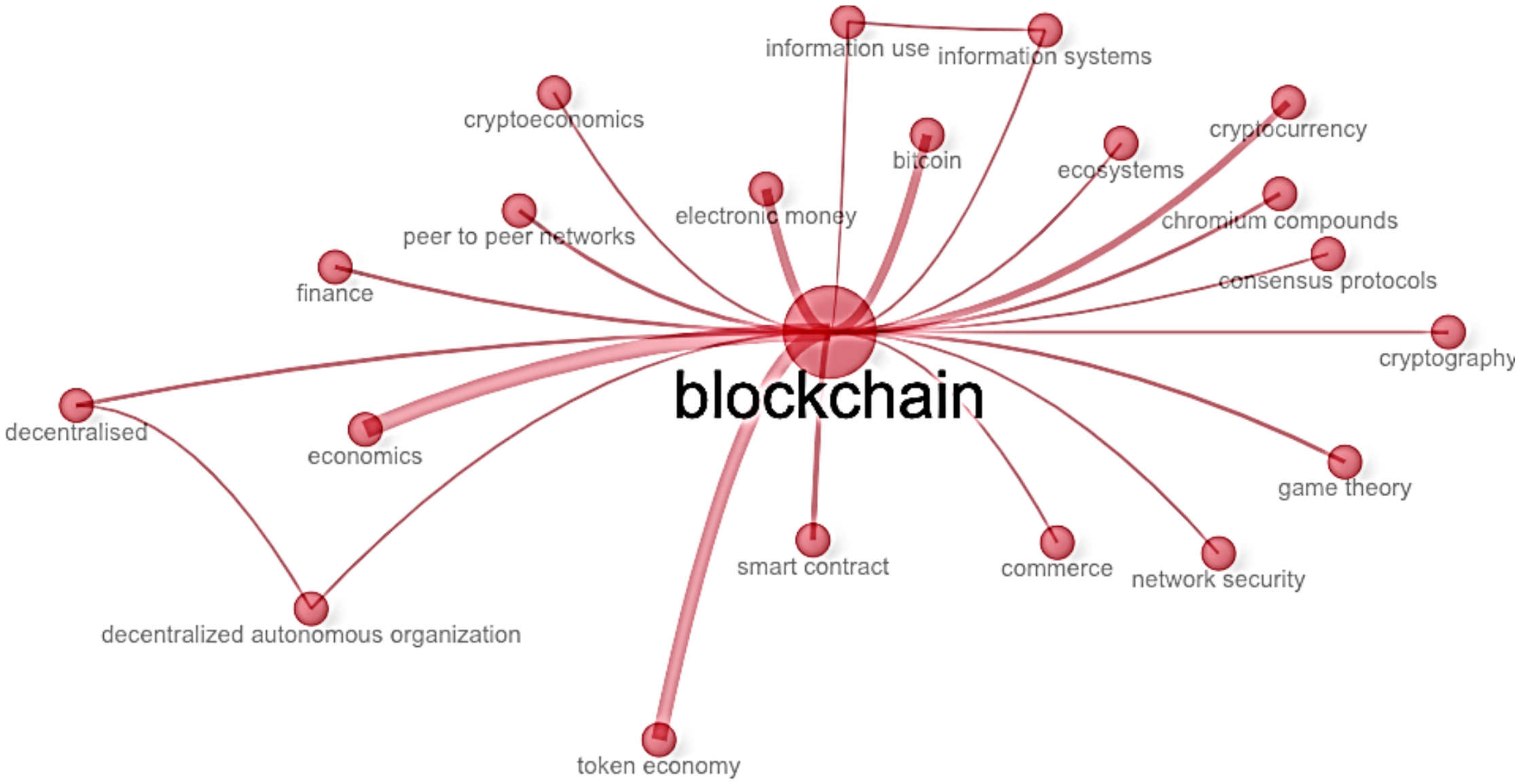

Fig. 5. Keyword Plus co-occurrence network

We selected words with at least 5 edges and obtained 21 words. Blockchain, economics, smart contract, token economy, and decentralization are among the most frequently repeated words in Fig 5. above, each associated with multiple concepts.

Unlike the repeated words in other networks, we observed that ten words were distinct in the keyword plus network and the disparate concepts are listed in Table 1. Among these concepts, Decentralized Autonomous Organizations (DAO)[50 51], cryptography[52], game theory[53], Peer-to-Peer[54], ecosystems[55], and information systems[56] are prominent in the field of cryptoeconomics. In addition, finance[57] is only present in the Keyword Plus and Abstract Co-occurrence networks. Unlike other co-occurrence networks, keyword plus is used to uncover emerging or essential terms in a field and identify hidden relationships between concepts[58].

Table 1. Disparate Concepts.

| Keyword Plus | Author's Keyword | Title | Abstract |
|---|---|---|---|
| electronic money | ethereum | internet | digital |
| cryptography | tokenization | mechanism | market |
| ecosystems | initial coin offering | | information |
| game theory | money laundering | | analysis |

| | | |
|---|---|---|
| commerce | smart cities | potential |
| DAO | token engineering | challenges |
| peer to peer information systems | symbolic modeling | process |
| systems | tokenomic modeling | trading |
| information use | tokenomic modeling | |

Second, the author's keyword co-occurrence network represents the relationships between keywords that authors use in their publications. This network is built by connecting the author's keywords that co-occur within the same documents[59]. Fig. 6. visualizes 24 keywords that appear at least 3 times.

The author's keywords co-occurrence networks explore the core concepts of research articles [60] . When we examine Fig. 6., we see that, apart from blockchain, the terms cryptoeconomics, token economy, cryptocurrency, and smart contract are used in conjunction with multiple concepts. Some concepts not present in the Keyword Plus, Title, and Abstract Co-occurrence Network but found in the Author's Keyword Co-occurrence Network include ICO, token engineering, tokenomics modeling, and tokenization (Table 1). Considering these concepts, it can be observed that the author's keywords are focused on cryptoeconomic design (CED)[61]. On the design side, Token Engineering and Token Economics are two sub-disciplines that can be identified. Trent McConaghy[62] (2018) was the first person to come up with the concept of token engineering.

- Token engineering: This technique aims to tokenize the system most efficiently and securely by utilizing system theory and control theory methods in designing an autonomous mechanism.
- Token economics focuses on the economic aspects of token-based systems. It involves analyzing the supply and demand dynamics of tokens, token distribution mechanisms, token utility, and the broader economic implications of the ecosystem.

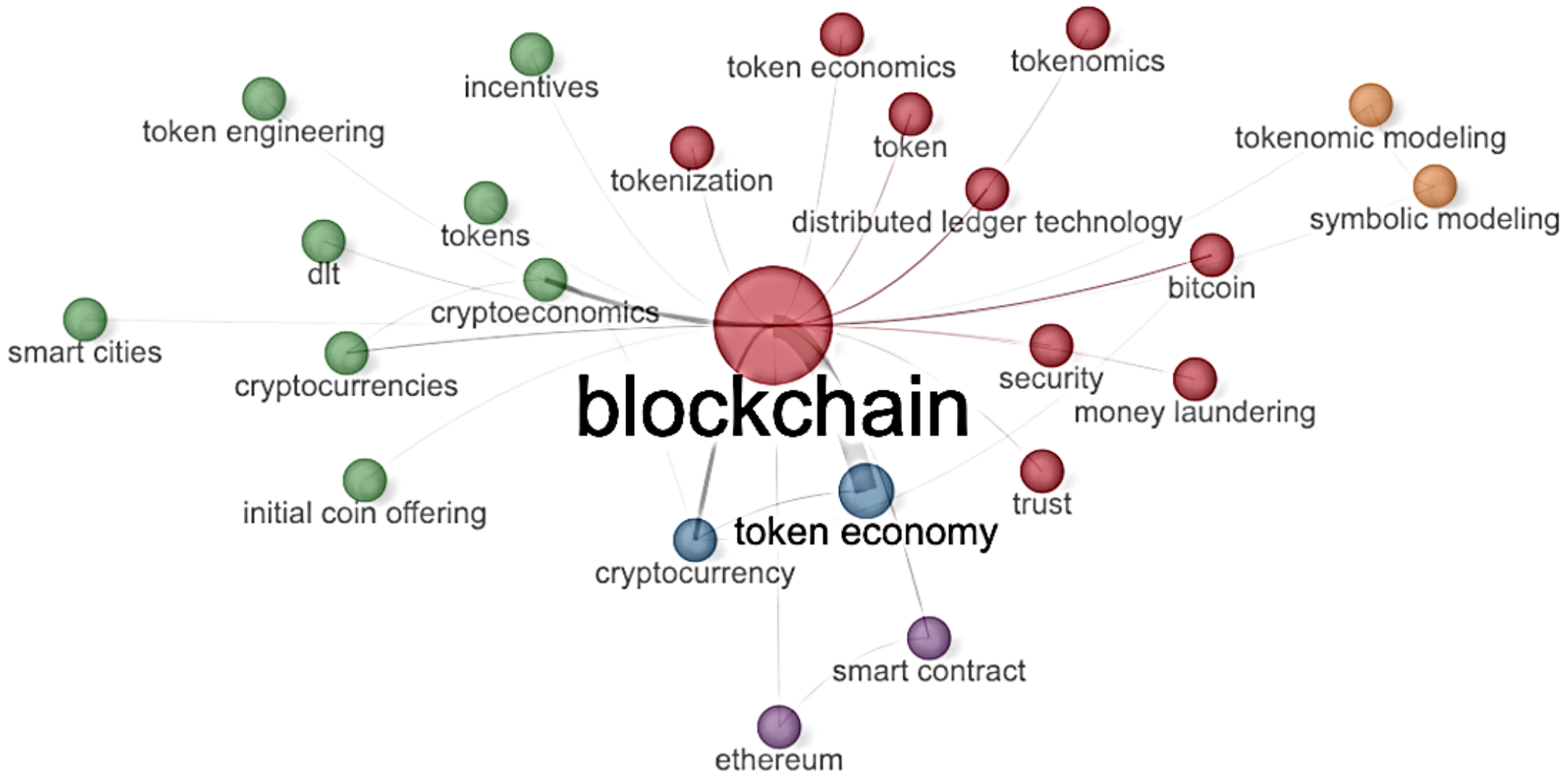

Fig. 6. Author's Keyword Co-occurrence Network

Third, In the title co-occurrence network, nodes represent paper titles, and edges connect titles when they co-occur together in the same documents. When we limited the words used in titles to those used together at least 3 times, we obtained 40 nodes. We displayed these nodes and their relationships in Fig. 7. Title co-occurrence networks can help identify common themes, topics, or trends across publications[63].

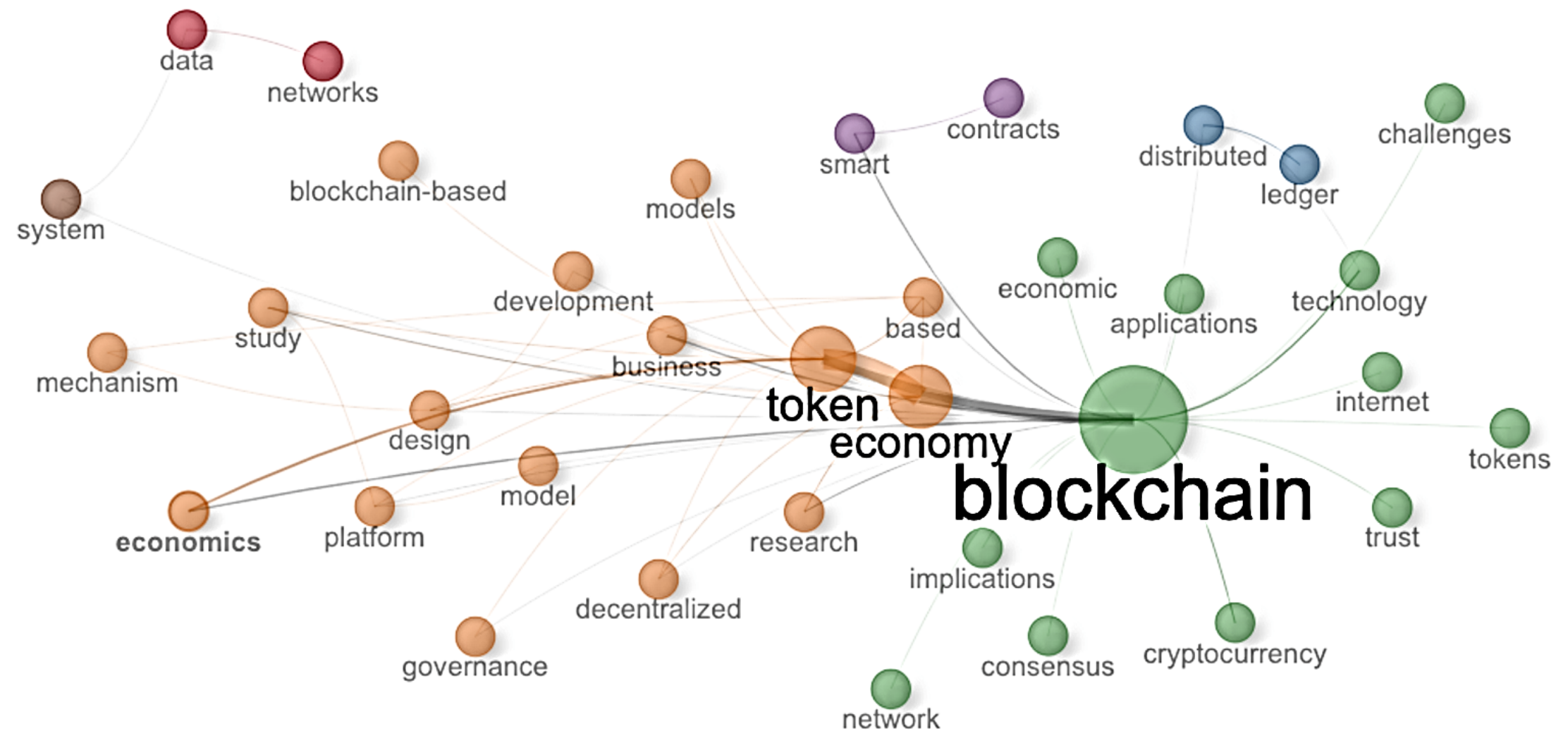

Fig. 7. Title Co-occurrence Network

We observe that concepts composed of two words, such as "smart contracts," "distributed ledger," and "token economy," cluster together. Along with "token economy," commonly used concepts like "decentralized," "design," "governance," "mechanism," "business," and "platform" are shown in the orange cluster. Additionally, "design" is the most commonly used word that appears together with other concepts. We can say that the orange cluster of the Title network (Fig 6.) is focused on the cryptoeconomic design side, as observed in the Author's keywords (Fig 5.). The green cluster led by blockchain is composed of more fundamental concepts such as "internet," "technology," "network," and "consensus."

In this co-occurrence network, only two concepts, "internet" and "mechanism," are present but not found in the other three networks (Table 1).

Lastly, abstract co-occurrence networks (Fig. 8.) can provide valuable insights into the connections and patterns among research topics or areas[64]. They can help identify clusters or communities of related topics and reveal the overall structure of the research landscape[65]. Nodes represent abstracts, and edges connect abstracts when they share common terms or phrases within text.

The Abstract Co-occurrence Networks (Fig. 8.) consist of 50 nodes divided into 3 clusters. The red cluster gathers around cryptoeconomics design (e.g., incentive, token, design, economy), the green cluster is related to fundamental concepts that make up Blockchain Technology (e.g., consensus, smart contract, trust, decentralized), and the blue cluster consists of essential concepts related to Blockchain technology (e.g., network, governance, system).

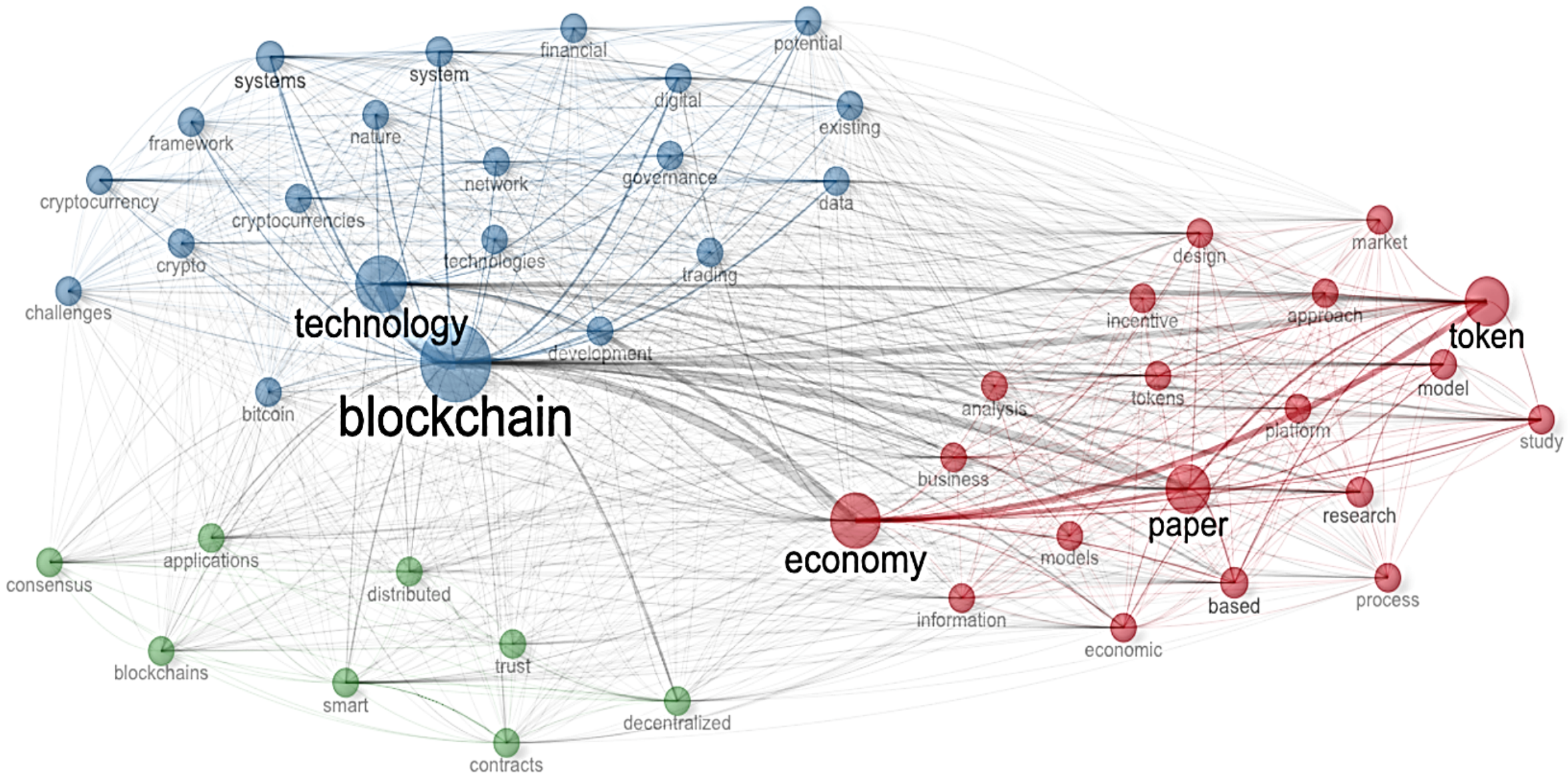

Fig. 8. Abstract Co-occurrence Network

Some of the concepts included only in the Abstract Co-occurrence Network are "market," "information," "digital," and "trading" (Table 1). Since abstracts summarize a scientific publication, Abstract co-occurrence networks provide a more detailed and content-rich analysis than other co-occurrence networks. Therefore, even though they remain outside the other co-occurrence networks, the concepts found only in Fig. 8 can offer different research perspectives.

Table 2. Common Concepts.

| Common Concepts | Freq. |
|---|---|
| Blockchain | 4 |
| Cryptoeconomics | 4 |
| Economics | 3 |
| Token economics | 4 |
| Smart contract | 4 |
| Bitcoin | 3 |
| Cryptocurrency | 4 |
| Distributed Ledger Technology | 4 |
| Decentralized | 3 |
| Token | 3 |
| Security | 2 |
| Consensus | 3 |
| Incentive | 2 |
| Trust | 3 |
| Network | 3 |
| Governance | 2 |

The conceptual structure we present by bringing these common concepts together provides researchers clear understanding of the key ideas in the literature and their significance.

## 4. Conclusion

This study delved into three fundamental structures that underpin cryptoeconomics research: Intellectual Structure, Social Structure, and Conceptual Structure. Visual representations of these structures were created through networks, providing researchers with accessible tools to navigate the intricate connections present in cryptoeconomics research.

The Intellectual Structure emerged through co-citation analysis. Frequently cited articles, influential authors, and the evolution of central themes were identified, offering a detailed map of the intellectual landscape of cryptoeconomics. The intellectual structure revealed through our analysis elucidated the multidisciplinary and interdisciplinary nature of cryptoeconomics research, shedding light on its institutional character. Additionally, it unveiled the theories underlying cryptoeconomics for institutional transformation. Also, a holistic understanding of cryptoeconomics' institutional and design aspects can assist system designers in better discerning opportunities.

The Social Structure was explored through co-author analysis, uncovering research collaboration patterns. The network analysis unveiled collaborative communities, highlighting the social dynamics facilitating innovation and knowledge exchange within the field. The collaboration network illustrated the convergence of diverse disciplines. However, international collaboration appeared relatively weak, possibly influenced by the nascent and evolving field. Enhancing international collaboration in future research endeavors would significantly contribute to the field's advancement.

The Conceptual Structure was elucidated by utilizing co-occurrence analysis. This revealed the shared keywords and concepts prevalent in article titles and abstracts, providing insight into the dominant themes and ideas within the field. Different co-occurrence networks have revealed hidden relationships alongside core concepts. However, due to the field's ongoing development, concepts tend to be used within a broader framework. As a result, there may be papers related to cryptoeconomics that we have not yet addressed. Future studies should consider this aspect and benefit from utilizing data from different databases.

This study, the first bibliometric analysis in cryptoeconomics as a research field, serves as a foundational resource, offering insights into the intellectual landscape, research productivity, and emerging trends. In this way, this paper contributes to a more profound understanding of cryptoeconomics. Moreover, it empowers researchers, policymakers, and industry sectors to make informed decisions, foster collaborations, and navigate the dynamic and evolving field of cryptoeconomics.

## Notes and References